\documentclass[twocolumn]{aastex62}
\usepackage{import}

\usepackage{mathtools}
\usepackage{chngcntr}
\usepackage{footnote}
\usepackage[caption=false]{subfig}
\usepackage{float}
\usepackage{tabulary,booktabs}

\usepackage{multirow}
\usepackage{graphicx}
\usepackage{amsmath}
\usepackage{yhmath}
\usepackage{mathtools}
\usepackage{csquotes}
\newcommand{\add}[1]{#1}

\def\zqual{\texttt{zqual}}

\usepackage{tikz}
\usepackage{tikz}
\usetikzlibrary{positioning}
\newdimen\nodeDist
\shorttitle{Galaxy–Environment Relations at Cosmic Noon}
\shortauthors{Chartab et al.}

\begin{document}

\title{\textbf{LATIS: Galaxy–Environment Relations at Cosmic Noon and the Role of Sample Selection}}

\correspondingauthor{Nima Chartab}
\email{nchartab@ipac.caltech.edu}

\author[0000-0003-3691-937X]{Nima Chartab}
\affiliation{Caltech/IPAC, 1200 E. California Blvd., Pasadena, CA 91125, USA}
\affiliation{The Observatories of the Carnegie Institution for Science, 813 Santa Barbara St., Pasadena, CA 91101, USA}

\author[0000-0001-7769-8660]{Andrew B. Newman}
\affiliation{The Observatories of the Carnegie Institution for Science, 813 Santa Barbara St., Pasadena, CA 91101, USA}

\author[0000-0002-8459-5413]{Gwen C. Rudie}
\affiliation{The Observatories of the Carnegie Institution for Science, 813 Santa Barbara St., Pasadena, CA 91101, USA}

\author[0000-0003-4218-3944]{Guillermo A. Blanc}
\affiliation{The Observatories of the Carnegie Institution for Science, 813 Santa Barbara St., Pasadena, CA 91101, USA}
\affiliation{Departamento de Astronom\'{i}a, Universidad de Chile, Camino del Observatorio 1515, Las Condes, Santiago, Chile}

\author[0000-0003-4727-4327]{Daniel D. Kelson}
\affiliation{The Observatories of the Carnegie Institution for Science, 813 Santa Barbara St., Pasadena, CA 91101, USA}

\author[0000-0001-7066-1240]{Mahdi Qezlou}
\affiliation{The University of Texas at Austin, 2515 Speedway Boulevard, Stop C1400, Austin, Texas 78712, USA}

\author[0000-0001-5803-5490]{Simeon Bird}
\affiliation{Department of Physics and Astronomy, University of California Riverside, 900 University Avenue, Riverside, CA 92521, USA}

\author[0000-0002-1428-7036]{Brian C. Lemaux}
\affiliation{Gemini Observatory, NSF NOIRLab, 670 N. A’ohoku Place, Hilo,
Hawai’i, 96720, USA}
\affiliation{Department of Physics and Astronomy, University of California, Davis, One Shields Avenue, Davis, CA 95616, USA}

\author[0000-0002-9336-7551]{Olga Cucciati}
\affiliation{INAF - Osservatorio di Astrofisica e Scienza dello Spazio di Bologna, via Gobetti 93/3, 40129 Bologna, Italy}

\begin{abstract}
\label{abstract}
We investigate the environmental dependence of galaxy properties at $z\sim2.5$ using the Ly$\alpha$ Tomography IMACS Survey (LATIS), which provides high-resolution three-dimensional maps of intergalactic medium (IGM) overdensity via Ly$\alpha$ forest tomography. Our analysis focuses on a UV-selected spectroscopic sample of 2185 galaxies from LATIS and a complementary set of 1157 galaxies from heterogeneous spectroscopic surveys in the COSMOS field. We compare these datasets to forward-modeled mock catalogs constructed from the IllustrisTNG300-1 simulation, incorporating realistic selection functions to match both LATIS and the literature sample. While the mass-complete simulation predicts strong environmental trends—more massive and quiescent galaxies preferentially occupy overdense regions—we find that such trends are significantly weaker or absent in the observed samples. The LATIS galaxies show no measurable correlation between specific star formation rate (sSFR) and IGM overdensity, a result reproduced by LATIS-like mock catalogs, confirming that UV selection systematically excludes passive and dusty galaxies in dense environments. The literature compilation, despite improved high-mass coverage, remains incomplete and affected by similar biases. We also analyze a mass-complete photometric sample from the COSMOS-Web catalog at $z\sim2.5$ and find no detectable sSFR–environment relation, a null result that our simulations indicate can be explained by photometric redshift uncertainties. In particular, we find no evidence for a reversal of the sSFR–density relation at cosmic noon. These results demonstrate that observed correlations can be heavily shaped by selection effects, and caution against inferring physical trends from incomplete spectroscopic samples. Deeper, more representative spectroscopic surveys are needed to robustly characterize environmental effects at this epoch.

\end{abstract}

\keywords{\small{Galaxy environments (2029); Galaxy evolution (594); High-redshift galaxies (734); Large-scale structure of the universe (902)}}

\section{Introduction}
\label{sec:Introduction}

Connecting galaxies to their large-scale environments at high redshift remains one of the most persistent challenges in observational astronomy. While it is well established that clusters and groups at $z<1$ are predominantly populated by quenched, early-type galaxies \citep[e.g.,][]{dres1980,Peng2010}, the influence of environment on galaxy properties at earlier epochs remains less certain. Early studies at $z\sim1$ reported a potential reversal of the local star formation–density relation, suggesting that galaxies in denser regions exhibited elevated star formation activity \citep[e.g.,][]{Elbaz07,Cooper08,Tran2010}. However, subsequent analyses incorporating more complete samples, and improved environmental metrics found no evidence for such a reversal at $z\sim1$ \citep[e.g.,][]{Patel2009,Quadri2012,Scoville13,darv2016,Kawinwanichakij2017,Tomczak2019,char2020,old2020}. 

These contrasting findings in the literature naturally raise the question of when and how environmental effects begin to emerge, and most critically, what the situation is during the peak epoch of galaxy assembly at cosmic noon ($z \sim 2$–3). This is a pivotal phase when galaxies are rapidly forming stars, yet the structures they inhabit—such as protoclusters—are still in the process of assembling. \citet{lema2022} reported that the star formation–density relation is reversed at cosmic noon, identifying a positive correlation between star formation rate (SFR) and local overdensity at a given stellar mass, driven by an overabundance of starburst galaxies in the densest regions. On the other hand, \citet{char2020} found that the same relation at this epoch qualitatively resembles that at $z\sim 1$ — but with a significantly weaker amplitude. Additional observational evidence has pointed toward the early onset of environmental quenching, particularly among the most massive galaxies. Recent studies suggest that quenching is already underway in dense environments by $z \sim 2$. \citet{kiyota2024} found that protocluster candidates at this epoch exhibit elevated quiescent fractions, especially at high stellar mass, even though star-forming members largely follow the same SFR–mass relation as field galaxies. \citet{ito2023} reported a protocluster at $z = 2.77$ in COSMOS containing at least 14 massive quiescent galaxies—a quiescent fraction nearly three times higher than the field, indicative of early red sequence formation. There are also examples of even higher-redshift ($z > 3$) protoclusters that show signs of early quenching \citep[e.g.,][]{McConachie2022,Tanaka2024}. \add{A key distinction in interpreting the environmental dependence of average star-formation activity in galaxies is whether the environment acts primarily by changing the relative abundance of quiescent versus star-forming galaxies or by shifting the star-forming main sequence (SFMS) at fixed stellar mass. The former channel dominates at low redshift \citep[e.g.,][]{Peng2010} and is suggested at higher redshift as well, whereas environment-dependent SFMS shifts remain debated at cosmic noon, with most studies finding little or no dependence} \citep[e.g.,][]{Koyama2013,darv2016,Chartab2021,Sattari2021,Shi2024}. Together, these findings underscore the complexity of galaxy–environment interactions at cosmic noon and motivate the need to investigate environmental effects using alternative, physically motivated tracers of large-scale structure.

One of the major barriers to progress in the field has been the difficulty of measuring galaxy environments in a precise and unbiased way over large volumes at high redshifts. Most studies rely on galaxy overdensity estimates based on photometric redshift catalogs, as deep spectroscopic coverage is limited and often incomplete. However, photometric redshifts carry significant uncertainties, which can smear out real structures along the line of sight. To mitigate this, some studies have adopted probabilistic or weighted density estimators, where each galaxy’s contribution to the local density field is weighted by its redshift uncertainty \citep[e.g.,][]{darv2015,cucc2018,lema2018,char2020,Taamoli2024}. While these methods help incorporate uncertainty information, they are sensitive to biases when the redshift error distribution is population-dependent \citep[e.g.,][]{Quadri2010}. Care must be taken when interpreting the resulting measurements, especially when comparing across heterogeneous samples. Moreover, when photo-$z$ uncertainties become comparable to or larger than the physical scale of large-scale structures, even weighted estimators struggle to recover meaningful environmental information.

These limitations motivate the need for alternative approaches that can provide physically motivated and unbiased measurements of the underlying matter distribution, independent of galaxy sampling. At $z \sim 2$–3, the intergalactic medium (IGM) traces the underlying dark matter distribution on Mpc scales with high fidelity \citep[e.g.,][]{lee2016,Qezlou22}. As such, it serves as a natural tracer of the evolving cosmic web. IGM tomography, based on Ly$\alpha$ forest absorption in the spectra of background galaxies and quasars, enables three-dimensional mapping of the IGM density field with resolution comparable to that of traditional galaxy-based environment measures. Crucially, this method is independent of the galaxy population, avoiding biases introduced by spectroscopic selection, luminosity limits, or incomplete sampling. The Ly$\alpha$  Tomography IMACS Survey \citep[LATIS; ][]{newm2020} utilizes this technique to reconstruct the IGM over contiguous volumes at $2 \lesssim z \lesssim 3$, offering a uniquely direct and physically grounded probe of environment during the epoch when large-scale structure and galaxy evolution are most tightly coupled.

In this study, we utilized the LATIS as a novel framework to investigate the relationship between galaxy properties and environment at cosmic noon. LATIS enables high-resolution IGM tomography, but assigning environmental metrics to individual galaxies requires accurate three-dimensional positions, and thus spectroscopic redshifts. We make use of the extensive LATIS spectroscopic sample of star-forming galaxies that were homogeneously selected and observed across the tomographic fields. While this sample offers uniform spectroscopic quality and well-defined selection criteria, it is not representative of the full galaxy population: the rest-frame UV selection inherently excludes quiescent galaxies and underrepresents dusty star-forming systems. To complement the LATIS sample, we incorporate the spectroscopic redshift compilation in the COSMOS field from \citet{khostovan2025}, which combines data from numerous publicly released spectroscopic campaigns. While this catalog increases the sample size of our analysis, it is still incomplete in key galaxy populations, particularly quiescent systems and heavily dust-obscured star-forming galaxies, which are critical for probing the full range of environmental effects. The compilation is also highly heterogeneous in terms of selection criteria, survey depth, and targeting strategies. These limitations make it essential to model the effective selection function when interpreting environment-dependent trends in the observed galaxy population. To quantify how these selection effects might bias our measurements, we use the IllustrisTNG300-1 cosmological hydrodynamical simulation to forward-model both the LATIS and COSMOS samples. The simulation captures a wide range of galaxy types and environments at $z \sim 2$–3, allowing us to construct mock catalogs that emulate the selection properties of our datasets. This enables a systematic assessment of how sample incompleteness and population biases impact the observed relationships between galaxy properties and IGM-defined environment. 

The remainder of this paper is organized as follows. In Section~\ref{sec:Data}, we describe the LATIS galaxy sample and the complementary literature compilation, both of which provide spectroscopic redshifts for environment studies. We also introduce the IGM tomographic maps, the photometric redshift sample used to probe mass-complete trends, and the IllustrisTNG300-1 simulation used for forward modeling. In Section~\ref{sec:Results}, we present our main results on the environmental dependence of stellar mass and sSFR, and evaluate how these trends are shaped by spectroscopic selection functions. Section~\ref{sec:photoz_env} explores the feasibility of recovering environmental correlations using photometric redshifts. In Section~\ref{sec:discussion}, we discuss the implications of our findings, particularly the impact of selection effects and redshift uncertainties. Finally, we summarize our conclusions in Section~\ref{sec:summary}.

Throughout this work, we assume a flat $\Lambda$CDM cosmology with $H_0=67 \rm \ kms^{-1} Mpc^{-1}$, $\Omega_{m_{0}}=0.3$ and $\Omega_{\Lambda_{0}}=0.7$. All magnitudes are expressed in the AB system, and the physical parameters are measured assuming a \cite{chab2003} IMF.

\section{Data}
\label{sec:Data}
In this paper, we use observational data from LATIS in conjunction with the TNG300-1 simulation to study the effect of the environment on the properties of galaxies, including their stellar masses and SFRs.

\subsection{Observations}

LATIS is one of the largest spectroscopic surveys of faint high-redshift galaxies to date, conducted over five years (2018–2022) using the Inamori-Magellan Areal Camera and Spectrograph \citep[IMACS;][]{dres2011} on the Magellan Baade telescope. The survey targeted Lyman-break galaxies (LBGs) and QSOs in the CFHTLS D1/D4 and COSMOS/D2 fields, covering 1.65 deg$^2$ across twelve IMACS footprints. A total of 7408 spectra were obtained, yielding 5575 high-confidence redshifts, of which 4176 correspond to galaxies and QSOs at $z > 1.7$.

LATIS produced three-dimensional tomographic maps of the IGM using Ly$\alpha$ forest transmission in 3012 background sightlines. The maps trace the Ly$\alpha$ flux contrast $\delta_F = F/\langle F \rangle - 1$ across 469,008 spectral pixels, sampling the $z = 2.2$–2.8 IGM at 4~$h^{-1}$~cMpc resolution. The final Wiener-filtered maps were smoothed with a Gaussian kernel of $\sigma_{\rm sm} = 4$~$h^{-1}$~cMpc. Each voxel contains a normalized transmission value $\Delta_F=\delta_F/\sigma_{\mathrm{map}}$, where $\sigma_{\mathrm{map}}$ is the map standard deviation for each field. Details of the observing strategy, target selection, map construction, and validation are provided in \citet{newm2020,Newman2025,Newman2025b}. Note that higher values of $\Delta_F$ correspond to higher transmission and, on average, lower densities.

This work focuses on a subset of 2185 galaxies drawn from the LATIS spectroscopic catalog with $M_* \geq 10^9\,M_\odot$ that lie within the redshift range $z = 2.2$–2.8, corresponding to the region covered by the IGM tomographic maps. The galaxies were originally selected using $ugr$ color cuts or photometric redshifts with $23.0 < r < 24.8$, making the sample intrinsically UV-selected. To ensure reliable stellar population measurements and environmental associations, only galaxies with high-confidence spectroscopic redshifts (\zqual{}~3 or 4) and sufficient near-infrared photometric coverage were included. AGNs, QSOs, stars, and blended or confused sources were excluded. Stellar masses and SFRs were derived using spectral energy distribution fitting as described in \citet{char2024}, employing the C++ implementation of {\tt LePhare}, with \citet{bru03} stellar population synthesis models, a \citet{chab2003} IMF, and the \citet{cal00} dust attenuation law. Both exponentially declining and delayed star formation histories were considered, and redshifts were fixed to the LATIS spectroscopic values during the fitting.

Each galaxy’s environment was quantified using the IGM overdensity, $\Delta_F$, extracted from the LATIS tomographic map at the galaxy’s three-dimensional position. The corresponding $\Delta_F$ value was assigned by identifying the galaxy’s location within the map grid and evaluating the voxel value at that position.

In addition to the LATIS spectroscopic sample, we incorporated a supplementary set of galaxies from the literature, compiled in \citet{khostovan2025}, which aggregates redshifts from multiple publicly released spectroscopic surveys in the COSMOS field. From this compilation, we selected 1157 galaxies with secure spectroscopic redshifts (quality flag 3 or 4), $M_* \geq 10^9\,M_\odot$, and within the redshift range covered by the LATIS tomographic map. Among these, 370 sources overlap with the LATIS sample, allowing for a direct comparison of redshift measurements. The two redshift sets exhibit excellent agreement, with a scatter of $\sigma_{\mathrm{NMAD}} = 0.0006$ and a median offset of $\Delta z_{\mathrm{med}} = -0.0002$, confirming the robustness of the combined catalog (see also the LATIS data release paper for a similar comparison; Newman et al. {\it submitted}). Although this compilation extends the dynamic range of galaxy properties, particularly at the high-mass end, it is heterogeneous in selection and observing strategy, and is therefore used primarily for comparison and cross-validation throughout this work.
\begin{figure}[t]
    \centering
    \includegraphics[width=\linewidth]{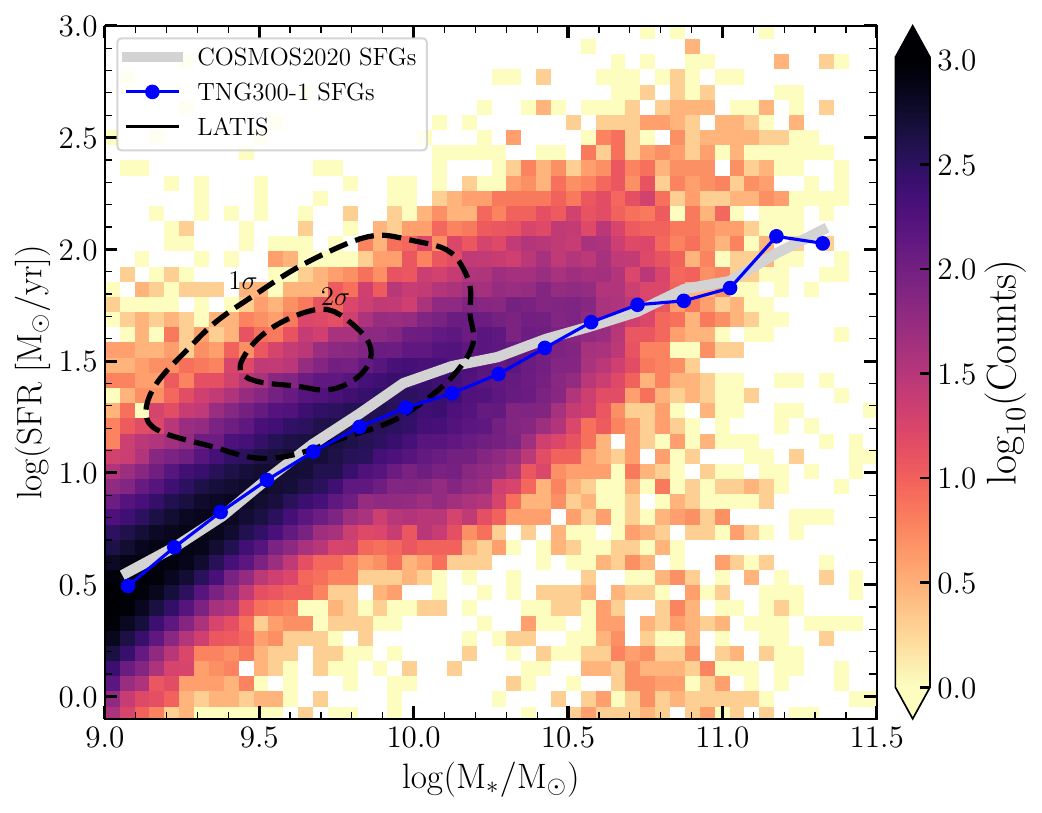}
    \caption{
    Comparison of the star-forming main sequence at $z \sim 2.5$ between simulations and observations. The heatmap shows the number density of TNG300-1 galaxies in the $\log(\mathrm{SFR})$–$\log(\rm M_*/M_\odot)$ plane. The light gray curve represents the empirical main sequence derived from COSMOS2020 star-forming galaxies. Blue points show the median SFR in stellar mass bins for TNG300-1 galaxies with $\mathrm{sSFR} > 10^{-10.5}\,\mathrm{yr}^{-1}$, after applying a +0.2 dex correction to the simulated SFRs. Dashed black contours show the LATIS spectroscopic galaxy sample, with the innermost and outermost contours approximating the 1$\sigma$ and 2$\sigma$ loci of the distribution.
    }
    \label{fig:mainseq_offset}
\end{figure}

\begin{figure*}[t]
    \centering
    \includegraphics[width=\linewidth]{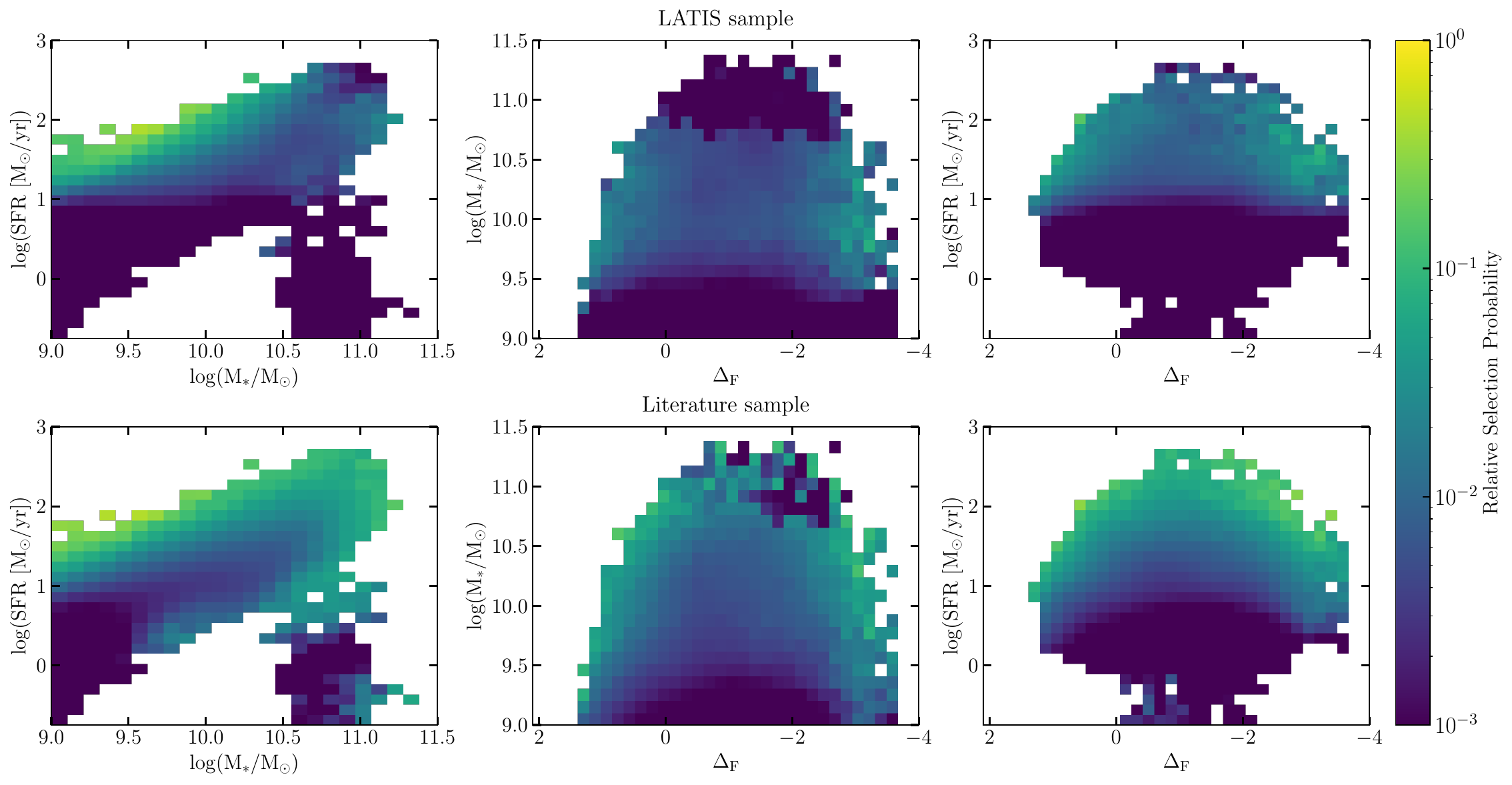}
    \caption{
Selection function maps for the LATIS (top row) and literature compilation (bottom row) spectroscopic samples. Each panel shows the median relative selection probability projected onto a two-dimensional space: stellar mass–SFR (left), $\Delta_F$–stellar mass (center), and $\Delta_F$–SFR (right). Darker regions indicate populations that are underrepresented due to selection biases. Both LATIS and the literature compilation systematically under-sample low-SFR galaxies, with LATIS further biased against massive galaxies, as expected for a UV-selected survey.
    }
    \label{fig:selection_function}
\end{figure*}

\begin{figure*}
    \centering
    \includegraphics[width=\linewidth]{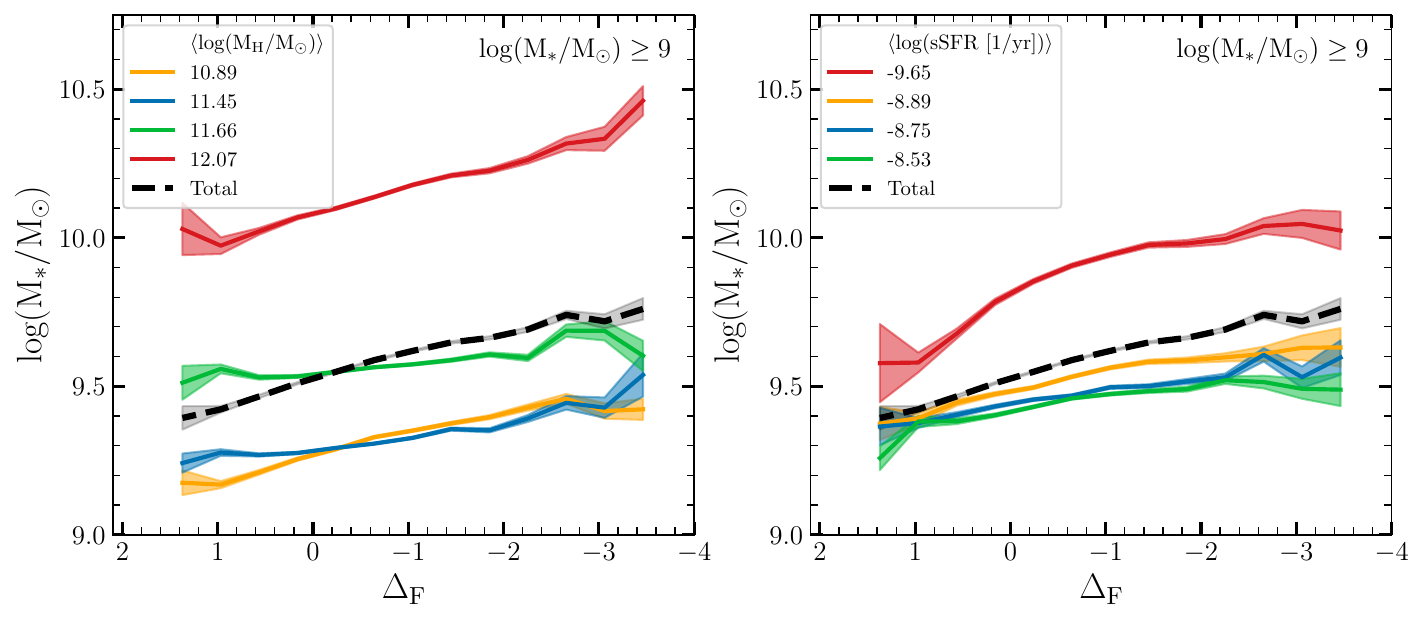}
    \caption{Stellar mass as a function of IGM overdensity $\Delta_F$ in the mass-complete TNG300-1 simulation ($M_* \geq 10^9\,M_\odot$). Left: Average stellar mass in bins of halo mass $M_H$. Right: Same, in bins of sSFR. In both cases, galaxies in overdense regions are significantly more massive on average.}
    \label{fig:tng_mass_env}
\end{figure*}

To ensure a consistent redshift scale across all datasets, we first applied velocity offsets of +190~km~s$^{-1}$ and +121~km~s$^{-1}$ to redshifts from VUDS \citep{LeFevre2015} and zCOSMOS \citep{lilly2007}, respectively, to bring them into alignment with the LATIS redshift scale, based on direct cross-matching of overlapping sources. Following the LATIS redshift calibration described in \cite{Newman2024}, we then applied a uniform +39~km~s$^{-1}$ shift to all redshifts, including LATIS, to place them on a systemic frame. These corrections ensure that all redshifts are referenced to a common systemic scale for subsequent analysis.

In addition to the spectroscopic datasets, we analyze a mass-complete sample of galaxies from the COSMOS-Web \citep{Casey2023} field with photometric redshifts from the COSMOS2025 catalog \citep{Shuntov2025} to assess the feasibility of detecting environmental trends using photo-$z$–based data. The methodology and results are presented in Section~\ref{sec:photoz_env}.

\subsection{Simulation}
To interpret our observational relations, we use the IllustrisTNG300-1 cosmological simulation and its simulated IGM absorption field. The IllustrisTNG simulations \citep{Marinacci2018, Naiman2018, Nelson2018, Pillepich2018, Springel2018} are a set of magnetohydrodynamical simulations performed with the AREPO code \citep{Springel2010}. IllustrisTNG300-1 is the largest simulated volume, covering a $(\sim300\,{\rm cMpc})^3$ box. The simulation follows the evolution of $2500^{3}$ dark matter particles and an equal number of baryonic resolution elements, with mass resolutions of $\sim 6\times10^{7}\,M_\odot$ and $1\times10^{7}\,M_\odot$ for dark matter and gas, respectively.

The simulation includes subgrid models for star formation, radiative cooling, chemical enrichment, black hole formation, and stellar and AGN feedback. These models are calibrated to match key low-redshift observables, including the galaxy stellar mass function, halo gas fractions, and the cosmic star formation rate density \citep{Weinberger2017, Pillepich2018}. Galaxy properties are well resolved for systems with $\gtrsim 100$ stellar particles \citep{Pillepich2018}; thus, we restrict our analysis to galaxies with $M_* \geq 10^9\,M_\odot$.

To extract galaxy properties, we analyze the $z = 2.58$ snapshot of TNG300-1. We adopt stellar masses within twice the stellar half-mass radius and the corresponding instantaneous SFRs. We verified that using 100 Myr–averaged SFRs does not change our results. In the simulation, a non-negligible fraction of galaxies exhibit ${\rm SFR} = 0$. While galaxies with functionally negligible star formation are physically expected at these epochs \citep[e.g.,][]{Kelson2014}, finite resolution in the simulation can also produce zero–SFR systems. To include these galaxies in log-space analyses while avoiding numerical divergence, we assign a floor value of ${\rm SFR} = 10^{-5} M_\odot\mathrm{yr}^{-1}$ \citep[see][]{Donnari2019}. This floor ensures that quiescent or low-SFR galaxies are retained in statistical measurements without artificially inflating their activity. In Section ~\ref{sec:discussion}, we explore the effect of varying the adopted SFR floor and confirm that our main results are insensitive to this choice.

Moreover, SFRs in hydrodynamical simulations often show systematic offsets from observationally derived values at fixed stellar mass, arising from differences in the treatment of feedback, star formation histories, and assumptions in SED modeling. To assess the consistency between simulated and observed galaxy populations, we compared the star-forming main sequence in TNG300-1 at $z \sim 2.5$ with that derived from the COSMOS2020 catalog \citep{Weaver2022} over the same redshift range. We find a median offset of $\sim$0.2 dex, with TNG300-1 galaxies exhibiting lower SFRs at fixed $M_*$. To account for this mismatch, we apply a uniform correction of +0.2 dex to the simulated SFRs. This adjustment brings the TNG300-1 star-forming sequence into agreement with the COSMOS2020 relation \citep{char2024}, as shown in Figure~\ref{fig:mainseq_offset}, and ensures consistency in subsequent analyses of selection bias and environmental trends.

We start with the idealized, noise-free tomographic IGM map of TNG300-1 from \citet{Qezlou22} to characterize large-scale environment. In this map, each gas particle is modeled as an absorber with internal properties smoothed using a quintic spline kernel, and the total absorption spectrum is constructed from the Voigt profiles of all particles along the line of sight. The neutral hydrogen fraction is computed assuming a uniform ultraviolet background, using the ionization equilibrium prescription described in \citet{Katz96}. To enable a consistent comparison with the reconstructed LATIS tomographic maps, we convert the simulated $\Delta_F$ values to their expected observational counterparts using the conditional mean relation $\langle \Delta_F \mid \Delta_F^{\rm true} \rangle$, as calibrated in Table 3 of Newman et al. {\it submitted}. Specifically, we apply the quadratic mapping
\begin{equation}
\left\langle \Delta_F^{\rm rec} \middle| \Delta_F^{\rm true} \right\rangle = c_0 x^2 + c_1 x + c_2,
\end{equation}
where $x = \Delta_F^{\rm true}$, ``rec'' denotes values in the reconstructed IGM maps, and the coefficients are $c_0 = 0.003$, $c_1 = 0.733$, and $c_2 = 0.018$. This transformation accounts for systematic differences between the true and reconstructed IGM fields.

\begin{figure*}[t]
    \centering
    \includegraphics[width=\linewidth]{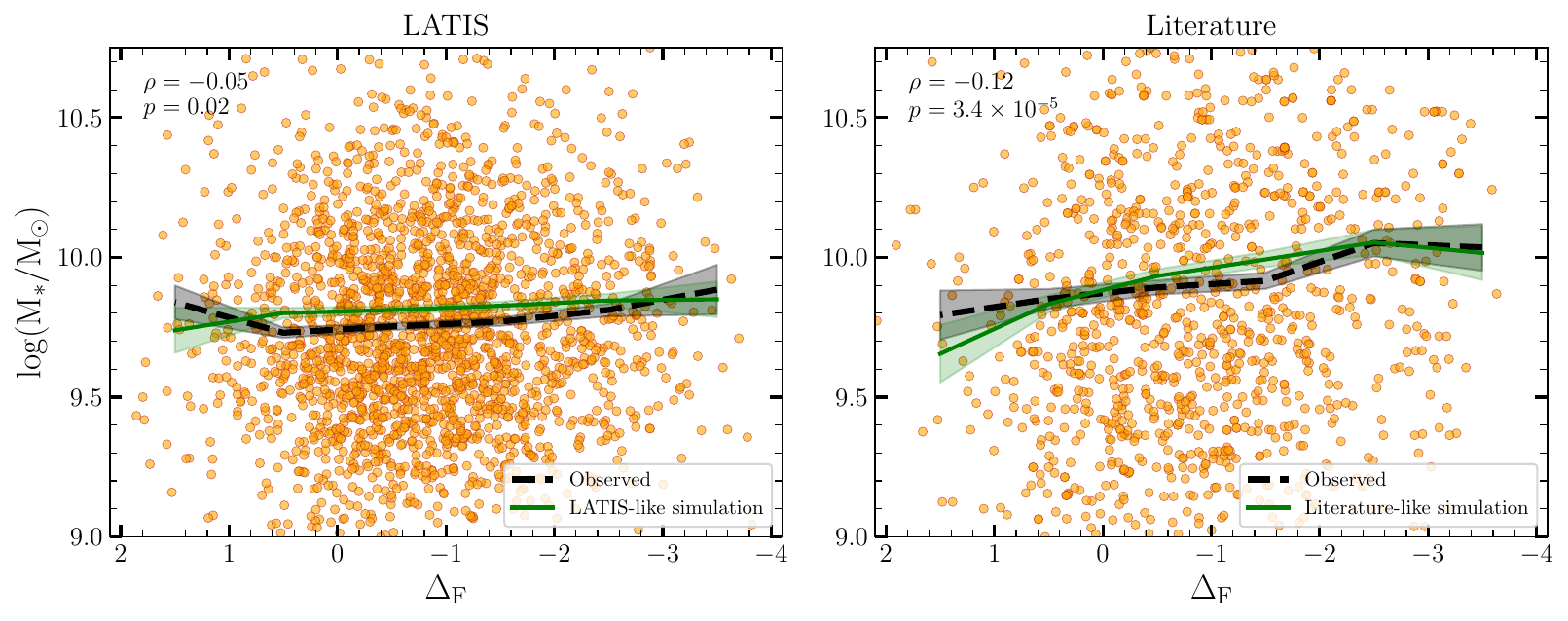}
    \caption{Stellar mass as a function of $\Delta_F$ in the LATIS (left) and literature (right) spectroscopic samples (orange points), compared to forward-modeled mock samples from TNG300-1 (green). Black dashed lines show the observed average trend. Spearman correlation coefficients and $p$-values are shown in each panel. LATIS exhibits a nearly flat $M_*$–$\Delta_F$ relation, whereas the literature sample shows a stronger correlation; selection-matched TNG mocks reproduce both, indicating that the observed amplitudes are primarily driven by selection effects.}
    \label{fig:latis_lit_mass_trend}
\end{figure*}

\subsection{Quantifying the Selection Function}
\label{sec:selection_function}

Spectroscopic galaxy samples are typically selected based on observational constraints and are rarely complete. In the case of LATIS, galaxies are selected based on their rest-frame UV, leading to an inherent bias against massive, dust-obscured star-forming galaxies and quiescent systems. To quantify how this selection bias may affect observed correlations with environment, we estimate the selection function as a function of stellar mass, star formation rate (SFR), and large-scale IGM overdensity, denoted by $\Delta_F$.

We fit three-dimensional kernel density estimates (KDEs) to both the observed and simulated galaxy populations in $(\log M_*, \log \mathrm{SFR}, \Delta_F)$ space. The KDE bandwidths are determined using \cite{scott1992}’s Rule and tested across a range of kernel widths to ensure robustness against oversmoothing or undersampling. For each simulated galaxy, we evaluate both the observed and TNG KDEs at its position and define a selection ratio as the pointwise ratio of the two densities. This provides an empirical estimate of the relative likelihood that a galaxy with given physical properties would be included in the spectroscopic sample. We normalize this selection ratio by its maximum value and assign it as a relative selection probability to each simulated galaxy.

Figure~\ref{fig:selection_function} shows the median relative selection probability in three two-dimensional projections: stellar mass–SFR, $\Delta_F$–mass, and $\Delta_F$–SFR. These maps reveal the underrepresentation of galaxies with low SFRs, and low/high stellar masses in the LATIS sample. This empirically derived selection function is later used to construct mock LATIS-like samples that reflect the observational biases of the survey.

\subsection{Mock Spectroscopic Samples}
\label{sec:mock_sample}

To assess how spectroscopic selection biases influence observed correlations with environment, we construct mock galaxy samples from the TNG300-1 simulation that replicate the selection functions of the LATIS and literature spectroscopic datasets. These LATIS-like and Literature-like catalogs allow direct, bias-aware comparisons with the underlying complete simulation, enabling us to isolate the impact of observational incompleteness.

Using the three-dimensional selection functions derived in Section~\ref{sec:selection_function}, each simulated galaxy is assigned a relative selection probability based on its stellar mass, SFR, and IGM overdensity ($\Delta_F$). We then draw galaxies from the simulation with replacement, using these selection probabilities as weights. The number of simulated galaxies is matched to the size of each observed sample. We verified that the resulting mock samples accurately reproduce the marginal distributions of the observed datasets in stellar mass, SFR, and $\Delta_F$\add{, and that truncating the lowest-probability tail of the selection function (i.e., the bottom 1\% of the cumulative probability distribution) does not affect our results.}

\section{Results}
\label{sec:Results}

\subsection{Environmental Dependence of Stellar Mass}
\label{sec:mass_env}

In hierarchical structure formation, dense environments are expected to host more massive galaxies due to earlier halo collapse, enhanced gas accretion, and higher merger rates \citep[e.g.,][]{White1978, Kauffmann1993}. This leads to an anticipated correlation between stellar mass and large-scale environment. We begin by examining this trend in the TNG300-1 simulation using a mass-complete sample of galaxies with $M_* \geq 10^9\,M_\odot$.

Figure~\ref{fig:tng_mass_env} shows the average stellar mass as a function of $\Delta_F$, split into bins of halo mass (left) and sSFR (right). In both cases, a strong environmental dependence is evident: galaxies in overdense regions are systematically more massive than those in underdense environments. For example, galaxies residing in halos with $\log(M_{\rm halo}/M_\odot) \sim 12$ show a $\sim$0.5 dex increase in stellar mass across the full range of $\Delta_F$. This trend is preserved even when splitting by sSFR, with the most quiescent galaxies showing the steepest increase in stellar mass toward dense environments. These results demonstrate that, in the absence of selection effects, the expected stellar mass–environment correlation is a robust prediction of galaxy formation models.

\begin{figure*}[t]
    \centering
    \includegraphics[width=\linewidth]{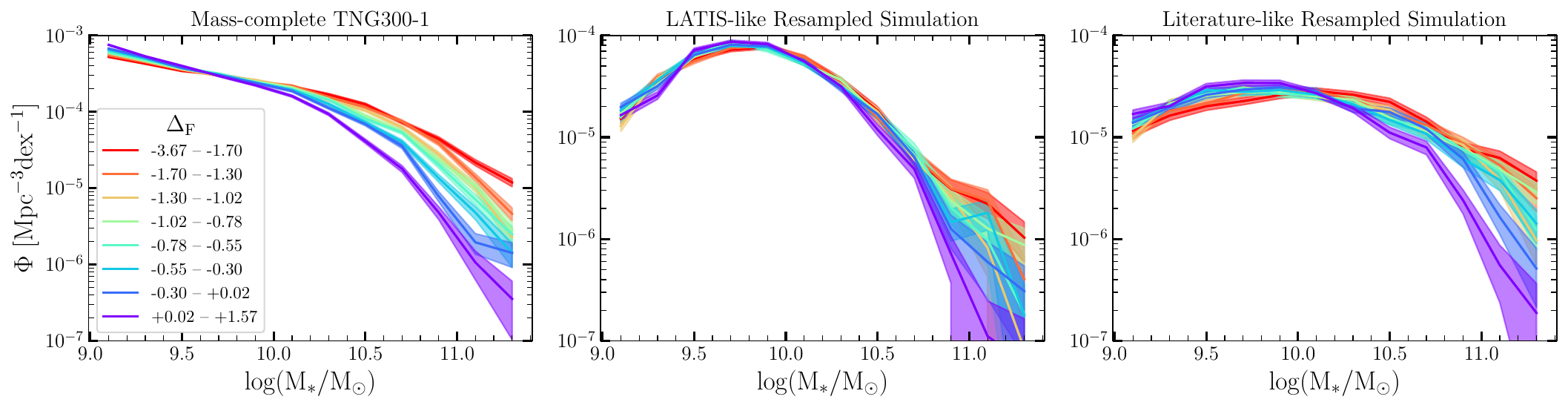}
    \caption{
    Environmental dependence of the stellar mass function (SMF) in the TNG300-1 simulation. Each panel shows the SMF across eight bins of IGM overdensity $\Delta_F$, with equal numbers of galaxies per bin. The left panel shows the mass-complete simulation, where overdense regions contain a higher number of massive galaxies than underdense regions. The center and right panels show LATIS-like and literature-like mock samples, respectively, where spectroscopic selection effects alter the observed environmental trends. Error bars for the mass-complete SMF represent Poisson uncertainties. For the LATIS-like and literature-like samples, uncertainties reflect the standard deviation over 500 bootstrap resamplings used to generate the selection-matched mocks. In the mass-complete TNG300-1 sample, overdense regions have a more top-heavy SMF; this intrinsic dependence is mostly suppressed in LATIS-like mocks and partially retained in literature-like mocks.
    }
    \label{fig:TNG_SMF}
\end{figure*}

\begin{figure*}[t]
\centering
\includegraphics[width=\linewidth]{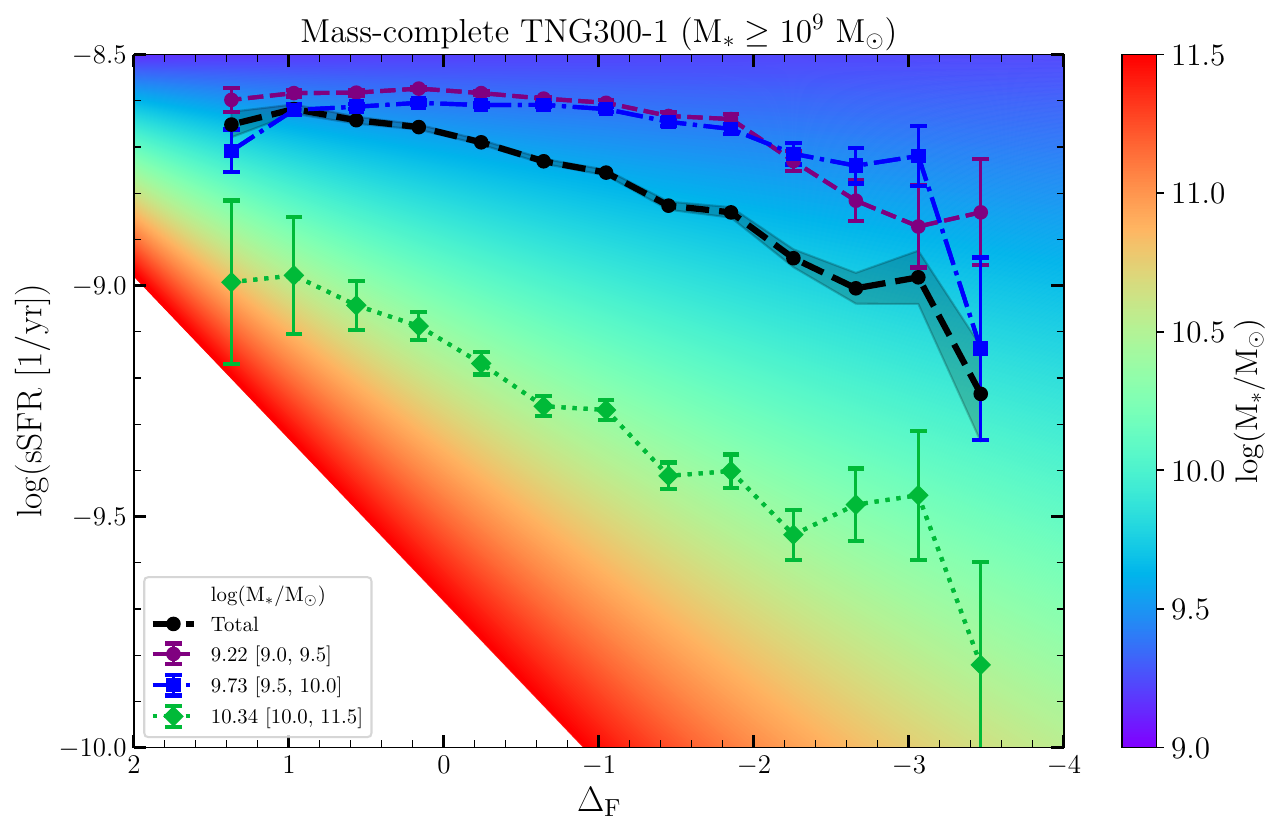}
\caption{sSFR as a function of $\Delta_F$ in bins of stellar mass, for the mass-complete TNG300-1 sample ($M_* \geq 10^9,M_\odot$). Points show the average $\log(\mathrm{sSFR})$ within each $\Delta_F$ bin for different mass intervals, with error bars indicating the error of the mean values.  The background color map shows the stellar mass $\rm \log(M_*/M_\odot)$ predicted from the best-fit model in Equation~\ref{eq:ssfr_model}, evaluated across the sSFR$-\Delta_F$ plane. A systematic decline in sSFR toward lower $\Delta_F$ is found in the simulation, particularly at higher stellar masses, indicating suppressed star formation in overdense regions.}
\label{fig:tng_fit}
\end{figure*}

Turning to the observational data, we examine the LATIS and literature spectroscopic samples in Figure~\ref{fig:latis_lit_mass_trend}. In LATIS, the trend is noticeably weak: the Spearman rank correlation between $\log(M_*)$ and $\Delta_F$ is $\rho = -0.05$ with a $p$-value of 0.02. While this result is marginally significant, the amplitude of the correlation is far lower than expected based on theoretical models. Following the methodology described in Section~\ref{sec:mock_sample}, we construct 500 forward-modeled LATIS-like mock samples by applying the empirically derived selection function to the TNG300-1 simulation. The green curve in Figure~\ref{fig:latis_lit_mass_trend} shows the average trend across these realizations. The forward-modeled mock closely reproduces the observed LATIS behavior, confirming that the lack of a strong mass–environment correlation is consistent with the effects of spectroscopic selection in the simulation. In particular, massive galaxies in overdense regions—often quiescent or dust-obscured—are missed in UV-selected samples like LATIS.

In contrast, the literature compilation shows a stronger correlation, with $\rho = -0.14$ and $p = 3.4 \times 10^{-5}$ (Figure~\ref{fig:latis_lit_mass_trend}, right). This may be due to its broader spectroscopic coverage and the inclusion of more massive galaxies. While still subject to selection effects—particularly at lower stellar masses and for quiescent systems—the literature sample provides better representation of the high-mass end, allowing the environmental dependence to emerge more clearly.

To summarize the effect of spectroscopic selection on the environmental dependence of stellar mass, we compute the stellar mass function (SMF) using three versions of the TNG300-1 simulation. These include the mass-complete sample, a LATIS-like mock, and a literature-like mock. Figure~\ref{fig:TNG_SMF} shows the SMF in bins of IGM overdensity $\Delta_F$, with each bin containing an equal number of galaxies. In the mass-complete case, a clear environmental trend is evident where overdense regions contain a greater number of massive galaxies than underdense ones, consistent with hierarchical structure formation. This trend disappears in the LATIS-like sample, consistent with the lack of a stellar mass–environment correlation in the observed LATIS data. In contrast, the literature-like sample retains a visible environmental trend, consistent with the pattern observed in the real literature compilation.

These results underscore that UV-selected spectroscopic samples, such as LATIS, are inherently limited in their ability to trace environmental trends in stellar mass. The galaxies most sensitive to environment—massive, quiescent, or dust-obscured systems—are systematically underrepresented, leading to flattened or suppressed correlations in observed data.

\begin{figure*}[t]
\centering
\includegraphics[width=\linewidth]{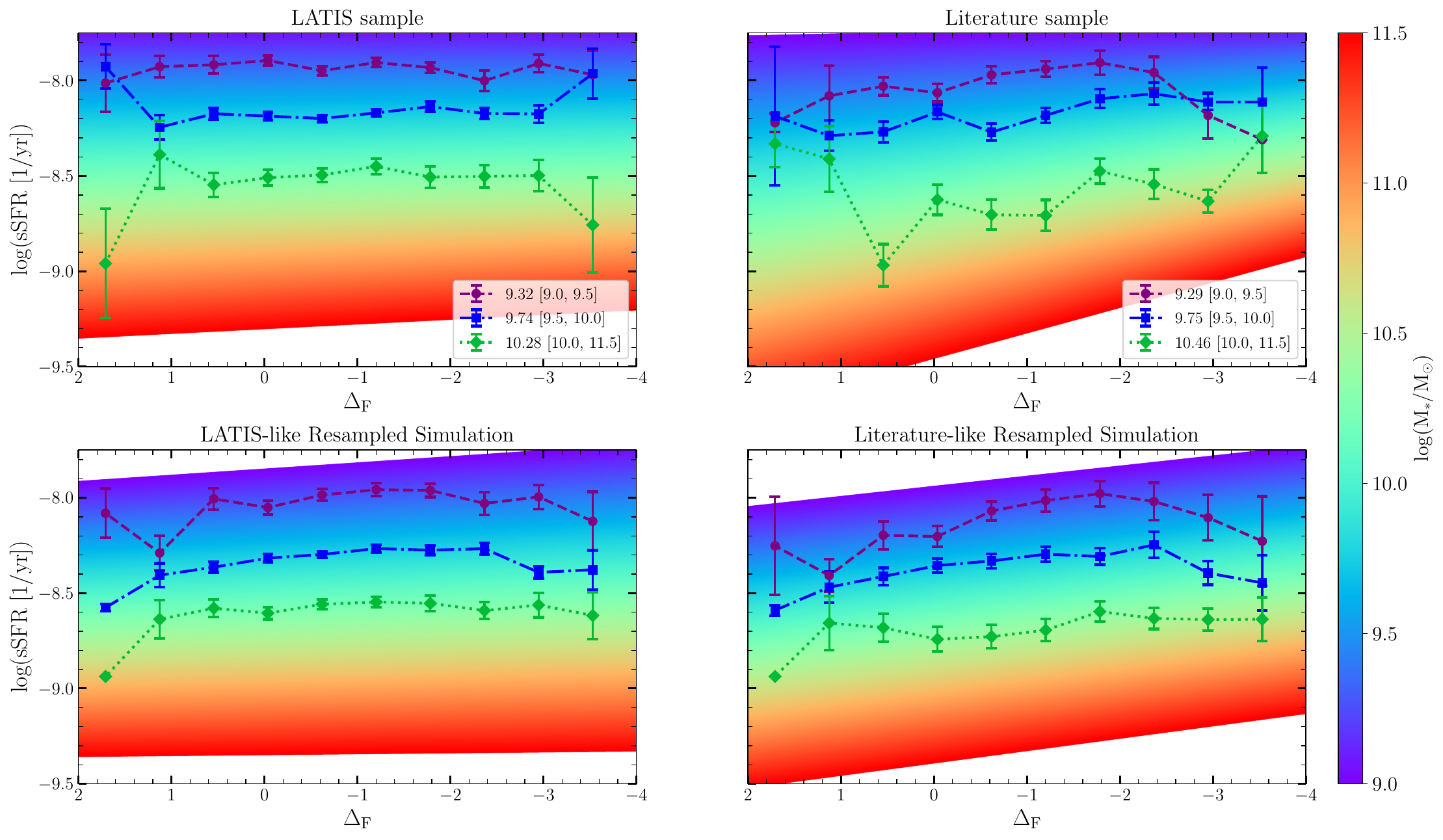}
\caption{Environmental trends in $\log(\mathrm{sSFR})$ as a function of stellar mass. Top: observational results from LATIS (left) and Literature \citep{khostovan2025} (right). Bottom: selection-matched mock samples from the TNG300-1 simulation. Background shows model predictions from the best-fit parameters. Points and error bars show the binned average log(sSFR) and uncertainties (standard deviation across 500 realizations in the mock samples). The declining sSFR–environment relation in the mass-complete TNG300-1 sample (Figure \ref{fig:tng_fit}) is flattened or weakly inverted once spectroscopic selection functions are applied, in agreement with LATIS and literature data.}
\label{fig:model_fits}
\end{figure*}

\subsection{A Parametric Model for Star Formation Activity}
\label{sec:fit_model}

To characterize the joint dependence of galaxy star formation activity on stellar mass and environment, we adopt a simple yet flexible parametric model for the specific star formation rate (sSFR). Specifically, we assume the following functional form:

\begin{equation}
\begin{split}
\log(\mathrm{sSFR}) &= \alpha + \beta \log\!\left(\frac{M_*}{10^{10}M_\odot}\right) + \gamma \Delta_F \\
&\quad + \eta \log\!\left(\frac{M_*}{10^{10}M_\odot}\right)\Delta_F
\end{split}
\label{eq:ssfr_model}
\end{equation}

\noindent where $\alpha$ is the normalization at $ M_*=10^{10}M_\odot$ and $\Delta_F = 0$, $\beta$ encodes the dependence of sSFR on stellar mass at fixed environment, $\gamma$ captures the linear sensitivity of sSFR to $\Delta_F$ at fixed stellar mass, and $\eta$ allows for a mass-dependent environmental response.

This form is motivated by the expectation that, to leading order, star formation activity may vary linearly with both galaxy mass and environment in log space. The inclusion of a mass–environment cross term accounts for the possibility that environmental effects are mass dependent—a trend supported by several studies at high redshift \citep[e.g.,][]{Balogh2016,darv2016,Kawinwanichakij2017,Pintos-Castro19,char2020}. This model is not intended to capture the full complexity of galaxy star formation histories, but rather to provide a practical framework for comparing environmental trends across different samples and selection functions.

We fit the model in Equation~\ref{eq:ssfr_model} to both the TNG300-1 simulation and the observed spectroscopic samples to investigate how galaxy star formation activity varies jointly with stellar mass and large-scale environment. The fitted model parameters are summarized in Table~\ref{tab:fit_params}, and we describe the key trends below.

In the mass-complete TNG300-1 sample ($M_*\geq 10^9 M_\odot$), the model reveals a clear negative dependence of sSFR on stellar mass and a mild but statistically significant sensitivity to environment. Both environmental terms ($\gamma$ and $\eta$) are positive, indicating that sSFR decreases in overdense regions and that this suppression becomes stronger at higher stellar mass.

Figure~\ref{fig:tng_fit} provides a visualization of the model behavior in the mass-complete TNG300-1 sample. It shows how average star formation activity varies with large-scale environment for galaxies in different stellar mass bins. The plotted points represent the average of $\log(\mathrm{sSFR})$ in bins of $\Delta_F$, with error bars indicating the standard error of the mean. A clear trend of increasing sSFR toward underdense regions (higher $\Delta_F$) is seen, especially among more massive galaxies. This behavior is captured by the positive environmental coefficients ($\gamma$ and $\eta$) in the best-fit model. The background color map, derived from Equation~\ref{eq:ssfr_model}, shows the corresponding stellar mass $\log(M_*/M_\odot)$ across the sSFR-$\Delta_F$ plane, illustrating how the model links mass, environment, and star formation activity across the full parameter space.

In contrast, the fits to the observed LATIS and literature spectroscopic samples yield markedly different trends. The stellar mass dependence ($\beta$) is steeper in both datasets compared to the simulation, while the environmental terms are consistent with zero or negative values. For LATIS, $\gamma$ and $\eta$ are statistically consistent with zero, indicating no measurable environmental modulation of sSFR. The literature sample exhibits mildly negative values, suggesting a weak enhancement of sSFR in overdense regions---opposite in sign to the trend in the mass-complete simulation.

To assess the role of selection effects, we apply the method described in Section~\ref{sec:mock_sample} to construct LATIS-like and literature-like mock samples by resampling the TNG300-1 simulation with selection probabilities derived in Section~\ref{sec:selection_function}. This procedure is repeated 500 times, and the model is fit to each realization. The mean values and standard deviations from these bootstrap fits are reported in Table~\ref{tab:fit_params}. For the LATIS-like mocks, the environmental coefficients remain close to zero, supporting the notion that UV-based selection in LATIS obscures intrinsic environmental trends. The literature-like mock preserves a weakly negative $\gamma$, qualitatively matching the observed sample.

Figure~\ref{fig:model_fits} presents the binned average log(sSFR) as a function of $\Delta_F$ in three stellar mass bins for the observed (top panels) and mock (bottom panels) samples. Colored backgrounds show model predictions using the best-fit parameters for each sample. Points and error bars show the running mean and associated uncertainty, computed from the observational data or from the dispersion among bootstraps in the mock panels. In the TNG300-1 mass-complete sample (Figure~\ref{fig:tng_fit}), sSFR declines in overdense regions for massive galaxies, a trend not recovered in either the observed or selection-matched mock samples. This discrepancy highlights the importance of selection effects: spectroscopic samples that omit quiescent and/or massive dusty galaxies systematically miss key environmental signals. 

\add{The decrease in average sSFR toward denser environments seen in the mass-complete TNG300-1 sample (Fig.~\ref{fig:tng_fit}) may reflect an increasing abundance of quiescent galaxies in dense regions and/or a shift of the SFRs of star-forming galaxies at fixed stellar mass (i.e., a shift of the SFMS). Although LATIS is not a fully representative census of all star-forming galaxies (Figure \ref{fig:selection_function}), its homogeneously selected star-forming sample shows no measurable sSFR–environment dependence, and therefore provides no evidence that environment shifts the SFMS locus. To assess this trend in the simulation, we repeated the fit to the full sample after excluding quiescent galaxies (requiring ${\rm sSFR}>0.2/t_{\rm univ}(z)$; \citealt{Pacifici16}, where $t_{\rm univ}(z)$ is the age of the universe at redshift $z$). For this star-forming subsample, we obtain $\gamma=0.002\pm0.001$ and $\eta=0.009\pm0.002$ (Table~\ref{tab:fit_params}), which are negligible, indicating that the predicted decline in the mass-complete case is driven predominantly by the changing balance between star-forming and quiescent galaxies, with any residual SFMS modulation by environment being at most very small. A spectroscopic sample that includes quiescent systems will be required to test this directly; we return to this point in Section ~\ref{sec:discussion}.
}

\subsection{Environmental Trends with Photometric Redshifts}
\label{sec:photoz_env}

\begin{table*}[]
\centering
\caption{Posterior Parameter Estimates for the sSFR--Mass--Environment Model}
\label{tab:fit_params}

\begin{tabular}{lcccc}
\hline
\hline
Sample & $\alpha$ & $\beta$ & $\gamma$ & $\eta$ \\
\hline
TNG300-1 (mass-complete)       & $-8.910 \pm 0.001$ & $-0.514 \pm 0.002$ & $0.117 \pm 0.001$ & $0.156 \pm 0.002$ \\
LATIS (observed)               & $-8.348 \pm 0.006$ & $-0.634 \pm 0.015$ & $-0.011 \pm 0.005$ & $-0.009 \pm 0.011$ \\
Literature (observed)          & $-8.430 \pm 0.008$ & $-0.683 \pm 0.015$ & $-0.060 \pm 0.006$ & $-0.049 \pm 0.011$ \\
TNG300-1 (LATIS-like mock)     & $-8.449 \pm 0.011$ & $-0.596 \pm 0.028$ & $-0.021 \pm 0.008$ & $0.012 \pm 0.020$ \\
TNG300-1 (Literature-like mock)& $-8.521 \pm 0.020$ & $-0.577 \pm 0.046$ & $-0.058 \pm 0.012$ & $-0.007 \pm 0.026$ \\
\add {TNG300-1 (star-forming sample)} & $-8.702 \pm 0.001$ & $-0.193 \pm 0.002$ & $0.002 \pm 0.001$ & $0.009 \pm 0.002$ \\
\hline
\end{tabular}
\end{table*}
The COSMOS-Web field, one of the key regions covered by LATIS tomographic maps, offers deep photometric coverage across a broad wavelength range, enabling environmental studies based on a more complete galaxy population. In particular, the recently released COSMOS-Web catalog \citep[COSMOS2025;][]{Shuntov2025} provides high-quality photometry from both ground- and space-based facilities, including JWST/NIRCam, and delivers robust photometric redshifts and physical parameters derived through SED fitting. At magnitudes of $m_{\rm F444W} \sim 26$~AB, the photometric redshifts exhibit a normalized median absolute deviation of $\sigma_{\rm NMAD} \approx 0.02$, an outlier fraction of 9\%, and a bias of 0.005, based on comparisons with spectroscopic redshifts from existing surveys. We use this photometric sample to investigate whether environmental trends in star formation and stellar mass—previously studied using spectroscopic data—can be recovered in a more complete population despite the presence of photometric redshift uncertainties.

To construct a representative sample, we select galaxies with photometric redshifts in the range $2.2 \leq z_{\rm phot} \leq 2.8$, F444W$< 26$ AB, and $\log(M_*/M_\odot) \geq 9$, excluding sources flagged as stars or affected by known quality issues such as the HSC star mask. To account for photometric redshift uncertainties, we generate 1000 Monte Carlo realizations by perturbing each galaxy's redshift according to a Gaussian distribution centered on the median photometric redshift (\texttt{zpdf\_med}) and with a standard deviation equal to half the width of the reported 68\% confidence interval. Given that the estimated bias is small compared to the measurement uncertainty, we do not apply any systematic correction. For each realization, galaxies are mapped into the LATIS tomographic map, and the corresponding $\Delta_F$ is extracted from the voxel nearest to each galaxy’s three-dimensional position. Galaxies falling outside the map boundaries are excluded on a realization-by-realization basis, resulting in an average of $\sim$8000 galaxies per realization.

\begin{figure*}[t]
    \centering
    \includegraphics[width=\textwidth]{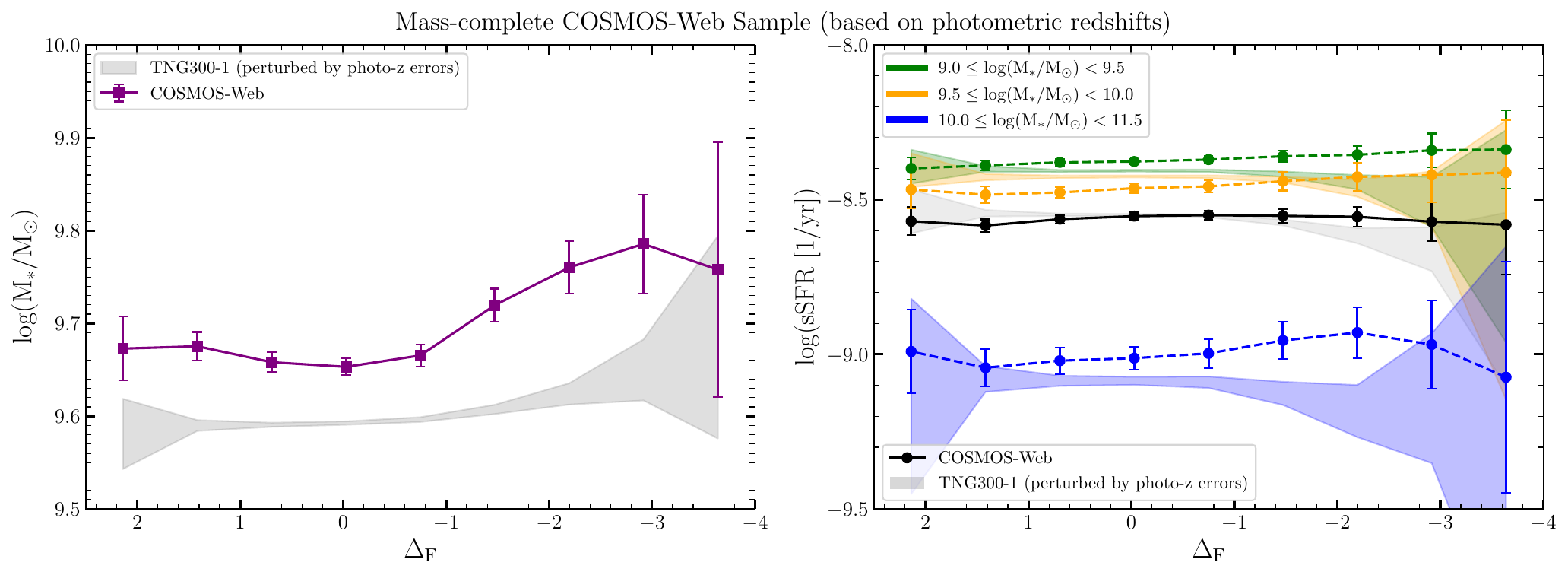}
    \caption{
    Environmental trends in stellar mass and sSFR for the mass-complete COSMOS-Web photometric sample at $z \sim 2.5$.
    \textit{Left:} Mean $\log(M_*/M_\odot)$ as a function of $\Delta_F$.
    \textit{Right:} Mean $\log(\mathrm{sSFR})$ versus $\Delta_F$ for the total sample (black line) and in three stellar mass bins.
    Error bars reflect the standard deviation across 1000 Monte Carlo realizations that incorporate photometric redshift uncertainties. In both panels, the shaded regions indicate the $\pm1\sigma$ scatter of the corresponding trends in the TNG300-1 simulation across Monte Carlo realizations in which galaxy redshifts are perturbed with the median COSMOS-Web photo-$z$ error, demonstrating that such uncertainties substantially wash out intrinsic correlations. Simulation-based sSFR have been offset by +0.2 dex to visually account for systematic differences in the measurement of physical parameters between the observational and simulated datasets.
    }
    \label{fig:cosmosweb_mass_ssfr_vs_dF}
\end{figure*}

We compute the mean $\log(\mathrm{sSFR})$ as a function of $\Delta_F$ in three stellar mass bins for each realization and also measure the mean stellar mass trend with environment. The final measurements and associated uncertainties are obtained by averaging over the 1000 realizations. As shown in Figure~\ref{fig:cosmosweb_mass_ssfr_vs_dF}, we find no significant correlation between sSFR and $\Delta_F$. However, a clear increase in mean stellar mass with decreasing $\Delta_F$ is observed.

To assess whether the absence of an observed sSFR–environment trend is driven by photometric redshift uncertainties, we generate 1000 Monte Carlo realizations of the TNG300-1 galaxy sample by perturbing the redshift of each galaxy using a Gaussian distribution with standard deviation $\sigma_z = 0.02(1+z)$, matching the median COSMOS-Web photo-$z$ uncertainty. For the snapshot used in this work at $z=2.58$, this corresponds to a comoving line-of-sight distance uncertainty of $\sim55\,h^{-1}\,\mathrm{cMpc}$. Perturbations are applied with periodic wrapping across the simulation volume. For each realization we recompute the environmental trends, and in Figure \ref{fig:cosmosweb_mass_ssfr_vs_dF}, the shaded regions show the $\pm1\sigma$ scatter across the ensemble, quantifying the impact of photo-$z$ errors on the detectability of an intrinsic sSFR–environment relation. We find that the resulting trends in the simulation are consistent with a complete flattening of the original correlation between sSFR and $\Delta_F$, closely matching the observational result. This agreement suggests that line-of-sight smearing introduced by photo-$z$ uncertainty significantly degrades the detectability of environmental trends, even in high-quality datasets such as COSMOS-Web.

These findings emphasize the importance of deep, representative spectroscopic surveys for recovering intrinsic correlations between galaxy properties and environment. Alternatively, future progress may come from improved photometric redshift methodologies that incorporate machine learning \citep[e.g.,][]{Pasquet2019,chartab2023} or joint SED and clustering constraints \citep[e.g.,][]{Newman2008} to reduce redshift uncertainties and enable environmental measurements across a broader population of galaxies.

\section{Discussion}
\label{sec:discussion}

Recently, a few studies have used spectroscopic samples to suggest a reversal of the star formation–density relation at cosmic noon. For example, \citet{lema2022} reported a positive correlation between sSFR and environment in the COSMOS field. Their sample, assembled from multiple surveys with diverse selection criteria—similar to the literature-based COSMOS spectroscopic compilation used in our study—is incomplete, particularly for quiescent and massive dusty galaxies. Our forward-modeling results show that such incompleteness can alter the observed trends, in some cases weakening or even reversing the intrinsic environmental correlations predicted by mass-complete simulations. Given that the LATIS galaxies—selected and observed in a uniform manner—show no significant environmental dependence, as indicated both by the best-fit model parameters and by a Spearman rank correlation test between $\log(\mathrm{sSFR})$ and $\Delta_F$ ($\rho = 0.006$, $p = 0.768$), and that the underlying mass-complete simulation sample predicts a suppression of sSFR in overdense regions, the significant positive correlation reported in some previous studies may be driven by selection bias rather than reflecting a genuine reversal. These findings suggest that, at least at $z \sim 2$–3, there is no compelling evidence for a reversal in the sSFR–density relation. Whether such a reversal occurs at earlier epochs ($z>3$) remains uncertain.

\begin{figure*}
    \centering
    \includegraphics[width=\textwidth]{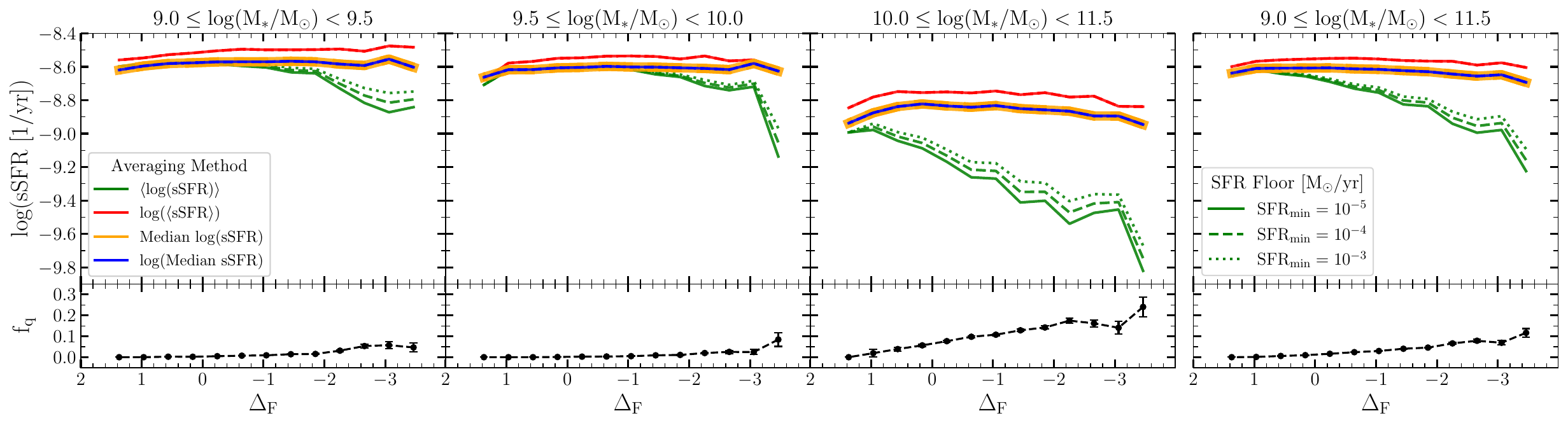}
    \caption{
    Environmental trends in sSFR and quiescent fraction from the mass-complete TNG300-1 simulation at $z = 2.5$. The top panels show $\log(\mathrm{sSFR})$ as a function of $\Delta_F$ in three stellar mass bins (left to center-right), with the rightmost panel showing the full galaxy sample. Colored curves represent four averaging methods commonly used in the literature: $\langle \log \mathrm{sSFR} \rangle$ (geometric mean; green), $\log \langle \mathrm{sSFR} \rangle$ (arithmetic mean; red), the median of $\log(\mathrm{sSFR})$ (orange), and $\log \mathrm{Median}(\mathrm{sSFR})$ (blue). Line styles denote different assumptions for the minimum allowed SFR ($10^{-5}$, $10^{-4}$, and $10^{-3}\ M_\odot\ \mathrm{yr}^{-1}$) for galaxies with unresolved or zero star formation. While all estimators exhibit consistent behavior for star-forming galaxies, only the geometric mean, $\langle \log(\mathrm{sSFR}) \rangle$, reliably tracks the growing quiescent population in overdense environments. This is particularly evident at high stellar mass, where the sSFR distribution becomes increasingly bimodal. The lower panels show the corresponding quiescent fraction ($f_q$) as a function of $\Delta_F$. Error bars reflect binomial uncertainty. The quiescent fraction increases steeply with environment, confirming that the dominant environmental effect in the simulation is to modulate the balance between star-forming and quiescent galaxies rather than shift the locus of the main sequence.
    }
    \label{fig:ssfr_env_methods}
\end{figure*}

At high redshift, the galaxy population exhibits a broad dynamic range in star formation activity, with a dominant star-forming sequence and a smaller, but non-negligible, population of quiescent systems. Moreover, extreme starburst systems are more common during this epoch, resulting in a highly skewed distribution of sSFRs. In such regimes, the choice of averaging statistic significantly affects the inferred environmental trends. Figure~\ref{fig:ssfr_env_methods} presents a comprehensive view of these effects using the mass-complete TNG300-1 simulation, showing sSFR as a function of $\Delta_F$ across stellar mass bins. Each panel displays trends computed using four commonly used averaging methods—$\langle \log(\mathrm{sSFR}) \rangle$, $\log(\langle \mathrm{sSFR} \rangle)$, median $\log(\mathrm{sSFR})$, and $\log(\mathrm{median\ sSFR})$—with line styles corresponding to three different SFR floor assumptions ($10^{-5}$, $10^{-4}$, and $10^{-3} M_\odot \mathrm{yr}^{-1}$) for galaxies with zero SFR in simulation. The mean of $\log(\mathrm{sSFR})$—equivalent to the logarithm of the geometric mean—offers a robust tracer of environmental effect, as it captures shifts in the distribution by minority population like quiescents. In contrast, $\log(\langle \mathrm{sSFR} \rangle)$, which represents the arithmetic mean in linear space, is disproportionately sensitive to highly star forming and starburst systems and can mask the contribution of quiescent galaxies. The median-based statistics—$\log(\mathrm{median\ sSFR})$ and median $\log(\mathrm{sSFR})$—track the central tendency of the star-forming sequence and are largely insensitive to the presence of quiescent galaxies unless they dominate the sample. As such, they are effective for identifying shifts in the main sequence but will miss trends in the quiescent fraction.

The lower panels of Figure~\ref{fig:ssfr_env_methods} illustrate this effect directly by plotting the quiescent fraction ($f_q$) versus environment. We use a redshift-dependent threshold of $\mathrm{sSFR} \leq 0.2 / t_{\mathrm{univ}}(z)$ \citep{Pacifici16} to define quiescent galaxies. In all stellar mass bins, we observe a strong, monotonic increase in $f_q$ toward lower $\Delta_F$ (i.e., toward denser environments), with the most rapid growth occurring at $\log(M_*/M_\odot) > 10$. This confirms that the primary environmental effect in the simulation is to modulate the fraction of quenched galaxies rather than to shift the locus of the star-forming sequence. Since $\langle \log(\mathrm{sSFR}) \rangle$ is sensitive to this shift in the overall population, it is the only estimator that tracks this trend reliably.

Moreover, these conclusions are robust to the assumed SFR floor value. While the absolute normalization of $\langle \log(\mathrm{sSFR}) \rangle$ shifts slightly depending on whether we adopt a floor of $10^{-5}$, $10^{-4}$, or $10^{-3} M_\odot\mathrm{yr}^{-1}$, the qualitative trend with $\Delta_F$ remains unchanged. This implies that the environmental dependence is not an artifact of how unresolved low-SFR galaxies are treated, but a genuine feature of the simulation’s galaxy population.

These findings provide a natural explanation for seemingly contradictory claims in the literature. Observations often report elevated quiescent fractions in dense environments at $z \sim 2$–3, and this trend is reproduced in the mass-complete TNG300-1 simulation. Statistics such as $\log(\langle \mathrm{sSFR} \rangle)$ or median-based measures can obscure environmental trends when the quiescent population remains a minority. This highlights that claims of elevated sSFR in overdense regions may arise not only from sample selection biases but also from the choice of averaging statistic. Interpreting such trends as evidence for an enhancement of star formation activity at fixed stellar mass in dense environments is not supported by our analysis. Instead, the predicted primary environmental effect is to modulate the quiescent fraction, and we find no indication of a true reversal of the star formation–density relation at fixed stellar mass during this epoch. 

Moreover, recent studies using photometric redshifts in the COSMOS field provide a useful point of comparison. \citet{Shi2024}, using COSMOS2020 and galaxy-count–based environmental estimates, find that sSFR shows no significant dependence on environment at $z>1$, consistent with our findings based on COSMOS-Web data and IGM-defined environments that account for photometric redshift uncertainties. However, \citet{Taamoli2024}, using similar photometric data, report that sSFR increases by $\sim 1$ dex from underdense to overdense regions at $z\sim 2-3$, for both the full and star-forming galaxy samples. These discrepancies may be partly due to uncertainties in environment reconstruction from photometric redshifts, which smooth out real structures along the line of sight and can bias the inferred environmental trends. Therefore a spectroscopic sample with high completeness is necessary for robustly characterizing the role of environment at this redshift.

\section{Summary}
\label{sec:summary}

We have studied the environmental dependence of galaxy properties at $z \sim 2.5$ using LATIS, which provides a physically motivated, three-dimensional measure of large-scale environment via Ly$\alpha$ forest tomography. Our analysis combines a UV-selected LATIS spectroscopic sample of 2185 galaxies with a supplementary set of 1157 spectroscopic redshifts compiled from heterogeneous surveys in the COSMOS field. To interpret these observations and assess the impact of sample incompleteness, we constructed forward-modeled mock catalogs based on the IllustrisTNG300-1 simulation, using empirically derived selection functions matched to the LATIS and literature samples. Our main findings are as follows:

\begin{enumerate}

\item The mass-complete simulation reveals strong environmental trends, with galaxies in overdense regions tending to be more massive and exhibiting suppressed sSFRs, particularly at the high-mass end. This supports a scenario where environmental quenching is already underway by $z \sim 2.5$ in the simulation.

\item The LATIS sample shows no measurable environmental dependence in either stellar mass or sSFR. Parametric modeling and Spearman correlation tests confirm the absence of a statistically significant trend.

\item In contrast, the literature compilation shows a reversed trend in sSFR—galaxies in overdense regions exhibit slightly elevated sSFRs relative to those in underdense regions. This apparent reversal is qualitatively reproduced in forward-modeled literature-like mocks, indicating that selection biases—particularly the underrepresentation of quiescent galaxies—can drive such trends.

\item Our forward-modeling approach demonstrates that UV- and emission-line–selected spectroscopic samples systematically miss the populations most sensitive to environmental effects in the simulation (e.g., quiescent or dust-obscured galaxies), flattening or even inverting the intrinsic relations seen in mass-complete simulations.

\item We analyze a mass-complete sample from COSMOS-Web using high-quality photometric redshifts and find no significant environmental trend in sSFR. This null result is consistent with expectations from simulation-based tests, which show that photometric redshift uncertainties at $z \sim 2.5$ effectively wash out intrinsic correlations.

\end{enumerate}
Taken together, our results suggest that observed correlations between star formation and environment at $z \sim 2.5$ are strongly shaped by selection effects. \add{Although the mass-complete TNG300-1 simulation predicts a modest suppression of sSFR in overdense regions at $z\sim2.5$, strongest for massive galaxies and driven primarily by a higher quiescent fraction rather than a shift of the star-forming main sequence, the observational datasets analyzed in this work are insensitive to this signal once selection effects and photometric-redshift uncertainties are accounted for. Accordingly, from these data alone we cannot determine whether an sSFR–density relation exists at this epoch, nor its sign. Robust tests will require more complete spectroscopy that includes quiescent and dusty systems alongside star-forming galaxies.}

\section*{Acknowledgments} 
\add{We thank the anonymous referee for providing insightful comments and suggestions that improved the quality of this work.} This paper includes data gathered with the 6.5 meter Magellan Telescopes located at Las Campanas Observatory, Chile. We thank the staff at Las Campanas Observatory for their dedication and support. N.C. and A.B.N. acknowledge support from the National Science Foundation under Grant No. 2108014.

\bibliography{LATIS_SFR}
\end{document}